\documentstyle[aps,prl,preprint,floats,epsfig,tighten]{revtex}
\newcommand{\nfig}[4]{\begin{figure}[htbp]\vfill\begin{center}
\mbox{\epsfig{figure=#1,width=#2}}\caption{#3}\label{#4}
\end{center}\vfill\end{figure}}
\title{Quarkonium Production at the Tevatron \\
 through Soft Colour Interactions\thanks{Work supported in part by the Swedish Natural Science Research Council and the U.S. Department of Energy, contract DE--AC03--76SF00515.}}
\author{A.~Edin$^a$, G.~Ingelman$^{a,b}$, J.~Rathsman$^c$}
\address{$^a$ Department of Radiation Sciences, 
Uppsala University, Box 535, S-751~21 Uppsala, Sweden\\
$^b$ Deutsches Elektronen-Synchrotron DESY, 
Notkestrasse 85, D-22603 Hamburg, Germany\\
$^c$ Stanford Linear Accelerator Center, 
Stanford University, Stanford, California 94309, USA}
\begin{document}
\preprint{\small{\vbox{\hfill SLAC--PUB--7506\\ 
					   \hspace*{1cm} \hfill TSL/ISV-97-0176\\
                	   \hspace*{1cm} \hfill May 1997\\}}}

\maketitle

\begin{abstract}
The direct charmonium and bottomonium production rate observed at
high-$p_\perp$ in $p\bar{p}$ collisions at the Tevatron is factors of
ten larger than predictions based on conventional perturbative QCD. 
We show that this excess can be accounted for by our model for 
soft colour interactions, previously
introduced to describe in a novel way the large rapidity gap events
observed at HERA.
\end{abstract}


Heavy quarkonia, i.e.\ bound states of heavy quark-antiquark pairs, are
thought to provide a useful testing ground for perturbative Quantum
Chromo Dynamics (pQCD). The reason being that the charm and bottom
quark masses provide a large scale which makes the short distance
behaviour calculable using a perturbative approach whereas the
nonperturbative contributions can be factorised into a wavefunction.
It was therefore a surprise that measurements by the 
CDF \cite{CDF}
and D{\O} \cite{D0} collaborations at the Fermilab Tevatron $p\bar{p}$ 
collider ($\sqrt{s}=1.8$ TeV)
gave cross-sections of direct high-$p_\perp$ 
$J/\psi , \psi^{\prime}$ and $\Upsilon , \Upsilon^{\prime}$ 
production far above the expectation from the colour singlet model
\cite{csm} based on conventional pQCD. The observed excess is 
generally an order of magnitude and increases to a factor $\sim 50$ 
with increasing $p_\perp$. 

In the colour singlet model pQCD is used to calculate the production of
a $Q\bar{Q}$ pair in a colour singlet state. This forms a quarkonium
state with the same angular momentum quantum numbers $^{2S+1}L_J$ by
coupling to the non-perturbative wave function at the origin, which is
obtained from the leptonic decay width. The striking failure of this
model has led to several phenomenological investigations and some new
models; for a review see \cite{braaten_fleming_yuan}. The colour octet
model, which is based on non-relativistic QCD, also takes
into account the more abundant perturbative production of $Q\bar{Q}$
pairs in colour octet states. The unknown probability for the
transformation into colour singlets due to non-perturbative processes
is parametrised with matrix-elements that have to be fitted to data,
but are universal so that the model can be tested by studying several
different processes. Similarly, in the colour evaporation model
\cite{cem} a certain fraction of all $Q\bar{Q}$ pairs, independently of
their production process and quantum numbers, form a quarkonium state.
Thus, both the colour octet model and the colour evaporation model
requires fitting to experimental data.
In contrast, our soft colour interaction (SCI) model \cite{sci}
gives a prediction also of the absolute rate which, as will be shown in
this Letter, is in good agreement with the Tevatron data. 

The SCI model was introduced as a novel way to
interprete the rapidity gap events observed in deep inelastic
scattering (DIS) at HERA. The model does indeed describe the salient
features of the ZEUS \cite{ZEUSgap} and H1 \cite{H1gap} data. The
conventional interpretation of these events is in terms of deep
inelastic diffractive hard scattering \cite{ingelman_schlein} on
partons in the pomeron, which is a colourless object exchanged from the
quasi-elastically scattered proton. Although phenomenological models
based on this idea work quite well to describe the
data, there are theoretical problems with the pomeron approach. 
In particular, the factorisation into a pomeron flux, a pomeron parton
density and a hard interaction described by a QCD matrix element may
not be universal for all processes, e.g.\ DIS and hadron collisions.
The pomeron-proton interaction is soft and thereby occurs on a long
space-time scale. It may therefore be incorrect to consider the pomeron
as decoupled from the proton during and after the hard scattering.

It is more natural to expect soft interactions with the proton both
before and after the snapshot of the DIS probe. To investigate this
line of thinking, we have developed a model \cite{sci} with a
mechanism for soft colour interactions as an alternative to the 
approach based on the pomeron and Regge phenomenology. 

The basic idea in our SCI model is that there may be additional soft
interactions, not previously accounted for, at a scale below the
cut-off $Q_0^2$ for perturbative QCD. Obviously, interactions do not
disappear below this cut-off. The question is how to take them into
account properly and whether conventional hadronization models give a
complete description. We propose that the quarks and gluons produced by
conventional pQCD processes, as described by matrix elements and parton
showers, interact non-perturbatively with the colour medium of the
proton remnant. This should be a natural part of the process in which
`bare' perturbative partons are `dressed' into non-perturbative ones
and the formation of the confining colour flux tube in between them.
These soft interactions cannot change the momenta of the partons
significantly, but they change their colour and thereby affect the
colour structure of the event. This in turn will lead to a modified
hadronic final state, e.g.\ with rapidity gaps or quarkonium. 

Lacking a proper understanding of non-perturbative QCD processes, we
have constructed a model \cite{sci} to describe and simulate such soft
colour interactions within conventional Monte Carlo (MC) event
generators. 
All partons from the hard interaction plus the remaining
partons in the proton remnant are assigned appropriate 
colour and anticolour charges. Partons from the hard interaction 
can make a soft interaction with partons in the proton 
remnant exchanging their colours but not changing
their momenta, which may be viewed as soft non-perturbative gluon
exchange. Similarly, partons in the proton remnant can exchange colour 
whereas no direct colour exchanges are allowed between the perturbative
partons. 
Since the process is non-perturbative the exchange
probability for a pair cannot be calculated from first principles and
is instead described by a phenomenological parameter $R$. The number of
soft exchanges will vary event-by-event and change the colour topology.
In the Lund string model \cite{lund} this corresponds to a modified
string stretching that alters the outcome of the hadronization. 

The SCI model was first implemented in {\sc Lepto} \cite{lepto}
simulating DIS. Here, the main effect producing rapidity gap events at
HERA is SCI between the partons from the hard scattering and those in
the proton remnant. In particular, with gluon initiated processes at
small Bjorken-$x$, the colour octet charge of the produced hard parton
system ($q\bar{q}$ plus possible additionally emitted partons) can turn
into a colour singlet. 
This may give a hadronic system $X$ separated in rapidity from
a very forward moving proton remnant system of small mass (e.g.\ a
proton or a resonance). The main features of diffractive scattering
emerges naturally and the HERA rapidity gap events can be surprisingly
well described by this simple model \cite{sci}. 

\begin{figure}
\vfill
\begin{center}
\mbox{
\epsfig{file=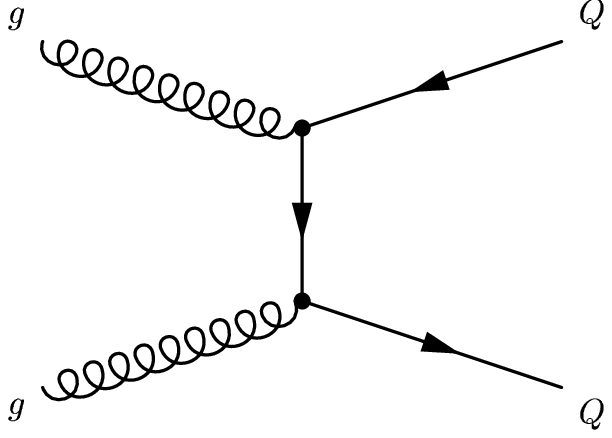,width=4cm}
\hspace{0.3cm}
\epsfig{file=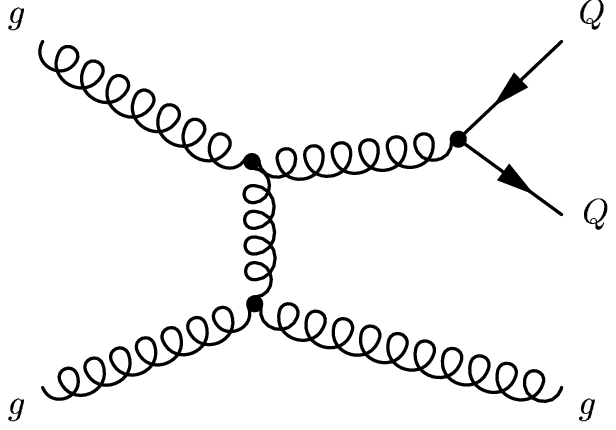,width=4cm}
}
\vspace{0.2cm}
\caption{Main pQCD processes for production of a $Q\bar{Q}$ pair
in (a) leading order and (b) next-to-leading order.}
\label{fig:feyn}
\end{center}
\vfill
\end{figure}

This SCI model can be straightforwardly applied to describe heavy
quarkonium production in $p\bar{p}$; a colour octet $Q\bar{Q}$ pair
from pQCD can be turned into a singlet and thereby form an onium state,
provided that the mass is appropriate. In leading order (LO) pQCD
heavy quark production occurs through $gg\to Q\bar{Q}$ (Fig.~\ref{fig:feyn}a)
and $q\bar{q}\to Q\bar{Q}$. 
However, higher order processes involving gluon splitting $g\to
Q\bar{Q}$ are important. For example, the next-to-leading order (NLO)
process $gg\to gQ\bar{Q}$ illustrated in Fig.~\ref{fig:feyn}b 
gives a large contribution because it is an $\alpha_s$ correction to
the large cross-section for gluon scattering ($gg\to gg$). Matrix
elements with explicit heavy quark masses are available up to NLO.
However, since the virtuality of the gluon need not be very large to
split, in particular to $c\bar{c}$, still higher orders may be
important at Tevatron energies. These can be taken into account
approximately through the parton shower approach. 

The LO and parton shower production of $Q\bar{Q}$ pairs are available
in the MC generator {\sc Pythia} \cite{pythia}. On the generated
parton level events we apply the above SCI mechanism which will turn
some $Q\bar{Q}$ pairs into colour singlets. A quarkonium state is then
produced if the invariant mass $m_{Q\bar{Q}}$ is below threshold for
open heavy flavour production ($2m_M$). 
Thus, the cross-section is 
\begin{equation}\label{eq:xsection}
\sigma_{onium} = \int_{2m_{Q}}^{2m_{M}}
  \frac{d\sigma^{{1}}}{dm_{Q\bar{Q}}}dm_{Q\bar{Q}} 
\end{equation}
where the singlet $Q\bar{Q}$ cross-section $d\sigma^1 = d\sigma
\otimes \mbox{SCI}$ is obtained from the application of the SCI model
on {\sc Pythia} events. This is not just a constant fraction of the
original $Q\bar{Q}$ cross-section, but depends somewhat on the
partonic state and the possible string configurations. Whereas the
number of possible string configurations increases with parton 
multiplicity, the number of string configurations giving a singlet
$Q\bar{Q}$ pair is more or less constant. Therefore, the relative 
probability for a singlet $Q\bar{Q}$ decreases slowly with parton
multiplicity. This in turn, implies a slight decrease 
($\sim$ 10~\% in the observed $p_\perp$-range) of the 
singlet $Q\bar{Q}$ fraction with increasing $p_\perp$ in the hard 
scattering process which causes more abundant parton showering.

The invariant mass of the $Q\bar{Q}$ pair is in principle given by the
pQCD process. However, we expect this mass to be smeared by the soft
colour interactions involving energy-momentum transfers of the order a
few hundred MeV. Irrespective of the original $Q\bar{Q}$ mass 
(below $2m_M$), we
therefore divide the quarkonium cross-section onto the different
quarkonium states based simply on spin-statistics as suggested by
\cite{cem}. This should be a good approximation since the
heavy quark system is nearly non-relativistic. The cross-section for a
given quarkonium state $X$ with total angular momentum $J_X$ is then given by
\begin{equation}
\sigma_X=\frac{\Gamma_X}{\sum_Y \Gamma_Y}\sigma_{onium}
\end{equation}
where $\Gamma_X=(2J_X+1)/n_X$
corresponds to a partial width. Here, we have included a suppression of
radially excited states, i.e.\ with the main quantum number $n_X$.  This
reproduces approximately the differences between different $\psi$ and
$\Upsilon$ states regarding their leptonic width and thus the wave function at
the origin \cite{buchmuller}, which should be particularly relevant when the
essentially pointlike $Q\bar{Q}$ pair forms an onium state.

The results of this model are shown in Fig.~\ref{fig:ptdep} and 
Table~\ref{tab2} together with the Tevatron data. The agreement is 
remarkably good, considering the simplicity of the SCI model and the
fact  that it was originally contructed for a different physics
issue.   As mentioned, the absolute normalization is {\em not} adjusted
to data,  but given by the model. The only parameter in the SCI model, 
i.e.\ the probability $R$ of a colour exchange between two partons,  is
kept at the value $R=0.5$ chosen to reproduce the rate of rapidity gap 
events in DIS at HERA. However, as for the rapidity gap rate, the
dependence of the onium rate on this parameter is quite small. For
example, changing  to $R=0.1$ only decreases the cross-section with 
$\sim 30$~\%.

\nfig{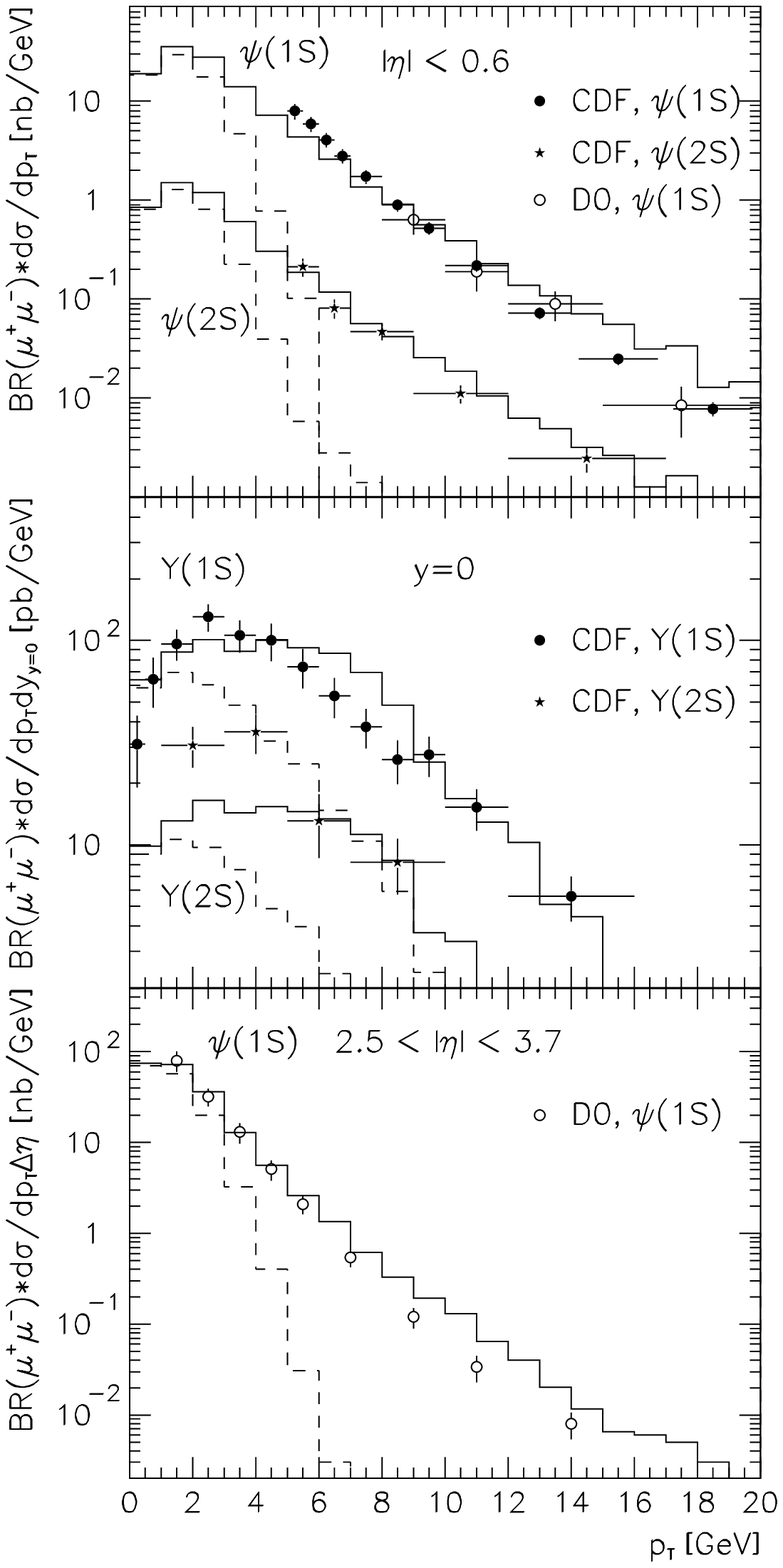}{90mm}{Transverse momentum 
spectra of prompt charmonium $J/\psi , \psi^{\prime} $ and bottomonium
$\Upsilon , \Upsilon^{\prime} $ in $p\bar{p}$ collisions at 
the Tevatron energy. Data from CDF \protect\cite{CDF}
(filled symbols) and D{\O} \protect\cite{D0} (open symbols) compared to 
results from the soft colour interaction (SCI) model applied to 
$c\bar{c}$/$b\bar{b}$ production from leading order ($\alpha_s^2$) 
matrix elements (dashed lines) and with the inclusion of higher order 
contributions calculated in the parton shower approach (full lines). 
Production central in pseudorapidity (a) $|\eta|<0.6$, (b) $|y|<0.4$ 
and forward (c) $2.5<|\eta|<3.7$ (where an assumed 20~\% systematic 
error has been added in quadrature).
 }{fig:ptdep}

There is however a strong dependence on the heavy quark mass $m_Q$. We
have used $m_c=1.35$ GeV and $m_b=4.8$ GeV. Increasing the quark
masses by $0.25$ GeV decreases the cross-section with a factor 3,
whereas a decrease of the quark masses by the same amount increases the
cross-section with a factor 2. This is as usual in pQCD heavy quark
production close to threshold and can be understood from 
Eq.~(\ref{eq:xsection}).

The cross-section also depends somewhat on other features of the Monte
Carlo model like the QCD $\Lambda$ parameter, details in the parton
shower model and the choice of parton density parametrizations. These
issues are not particular for the SCI model, but of a general
character. We therefore do not adjust any of these features, but leave
them as they are by default in the {\sc Pythia} MC model which provides
a standard for many processes and observables. We note that there are
no significant changes in our results when varying the virtuality scale
for the parton shower or using different available parton density
parametrizations (e.g.\ MRS or CTEQ) based on recent data. However,
including the multiple parton-parton interactions in {\sc Pythia}
lowers the total onium cross-section slightly ($\sim 20~\%$).

The observed $p_\perp$ spectra of the different onium states are also 
quite well reproduced by the model, as demonstrated in
Fig.~\ref{fig:ptdep}. The contribution from the LO process (shown
separately as dashed curves)  have a steeper $p_\perp$ dependence and
only contributes significantly at lower $p_\perp$, in particular for
$\psi , \psi^{\prime}$. Thus, the higher order contributions are most
important. 

Comparing model and data in detail, there is a tendency for the model
to have a slightly flatter $p_\perp$ distribution for the $J/\psi$
production. 
This may be an effect of using the approximate parton shower gluon
splitting $g\to c\bar{c}$ instead of the NLO matrix element which might
be more appropriate at large $p_\perp$. It may also be influenced by
the detailed contribution of different onium states, since $J/\psi$
from decays contribute more at smaller $p_\perp$. 

The relative rates of $J/\psi$ from direct production and from decays
of higher onium states are given in Table~\ref{tab2} together with
some ratios of cross-sections for different onium states. The model
gives good agreement with data in most cases. This shows that simple
spin statistics can describe the main effects. The only deviation of
some significance is the ratio 
$\sigma(\Upsilon(2$S$))/\sigma(\Upsilon(1$S$))$, where the simple 
suppression of radially excited states seems too strong.

\begin{table}[htb]
\caption{Relative rates in~\% \ for 
different onium states in the Monte Carlo model and the 
Tevatron data \protect\cite{CDF}. The charmonium cross-sections
are in general for $p_\perp \ge 5$ GeV and $|\eta|<0.6$ whereas
the bottomonium cross-sections are for $|y|<0.4$. } 
\label{tab2}
\begin{tabular}{ldd} 
                                & MC model             & Data       \\
\hline    
 $J/\psi$ direct                & 61                   & 64 $\pm$ 6  \\
 $J/\psi$ from $\chi_c$         & 26                   
 & 30 $\pm$ 6 $^{\dag}$ \\
 $J/\psi$ from $\psi^\prime$    & 13                   &  7 $\pm$ 2
   $\leftrightarrow$ 15 $\pm$ 5 $^{\ddag}$ \\
 $\sigma(\chi_c^2)/(\sigma(\chi_c^1)+\sigma(\chi_c^2))$   
                                & 62                   & 47 $\pm$ 9  \\
 $\sigma(\psi^\prime)/\sigma(J/\psi)$   
                                & 32                   & 25 $\pm$ 6 $^{\S}$ \\
 $\sigma(\Upsilon(2$S$))/\sigma(\Upsilon(1$S$))$   
                                & 29                   & 46 $\pm$ 9   
\end{tabular}
\small{
$^{\dag}$ Data are for $p_\perp \ge 4$ GeV.\\
$^{\ddag}$ Lower (upper) value for $p_\perp \sim$ 5 (18) GeV. \\
$^{\S}$ Assuming the same fraction of J/$\psi$ and $\psi^\prime$ from 
      $b$ hadron decays. }
\end{table}

The SCI model provides a mechanism for how a colour octet $Q\bar{Q}$
pair is turned into a singlet and thereby an absolute normalization of
the cross-section for the quarkonium production. In addition, one
obtains a direct relation to other manifestations of soft interactions,
such as diffraction and rapidity gaps. In the colour octet model the
probability to form a singlet is parametrized into matrix elements, but
no explicit mechanism for how an octet is turned into a singlet is given.

These models are in clear contrast to the colour singlet model for 
quarkonium production
which cannot explain the large rate observed at the Tevatron. 
In addition, they differ in the fragmentation function of $J/\psi$, 
i.e.\ $D_{g\rightarrow J/\psi}(z,\mu)$ where $z$ is the 
momentum fraction of the $J/\psi$ relative to the `mother' gluon
and $\mu$ is the factorisation scale.
In the colour singlet model the fragmentation
function is relatively flat \cite{braaten_yaun1}, whereas in the colour
octet model it is essentially a $\delta$-function at $z=1$ \cite{braaten_yaun2}.

The fragmentation function can be measured 
by taking the ratio of the transverse momentum of the $J/\psi$ and
the total transverse momentum in a cone centered at the $J/\psi$
\cite{EL_quarkonia}.
Following conventional jet algorithms for high energy
hadronic collisions, one may choose a cone $\Delta R = \sqrt{\Delta
\phi ^2 +\Delta \eta ^2} \simeq 0.7$ in pseudorapidity and azimuthal
angle. Applying this to our SCI model results in a fragmentation
function for $J/\psi$'s with $p_\perp>5$ GeV which is strongly peaked
at $z=1$. Although its exact form depends on details in the model, it
is essentially as expected in the colour octet model and very different
from the colour singlet model.

In summary, we have shown that the large rate of direct high-$p_\perp$
charmonium and bottomonium in high energy $p\bar{p}$ collisions can be
quite well described the Soft Colour Interaction model. The same SCI
model also accounts for the rapidity gap events observed at HERA. This
indicates that these features of non-perturbative strong interactions
are quite general and can be described by a universal model.





\end{document}